\documentclass[12pt,cite,epsf,epsfig]{article}
\usepackage{epsfig}
\title{Revisiting 5D chiral symmetry breaking and holographic QCD models}
\author{Namit Mahajan\thanks{E--mail : nmahajan@mri.ernet.in}\\
	{\em Harish-Chandra Research Institute,} \\
	 {\em Chhatnag Road, Jhunsi, Allahabad - 211019, India.}}

\def\be{\begin{equation}}
\def\ee{\end{equation}}
\def\bea{\begin{eqnarray}}
\def\eea{\end{eqnarray}}

\begin{document}
\maketitle

\begin{abstract}
We take a closer look at the recently discussed models of hadrons based on
holographic ideas and chiral symmetry breaking in five dimensions. We study the
two point correlator in detail and look at the field theoretic properties
needed to be satisfied by the correlator. It is shown that the spectral
density becomes negative, violating the
basic requirements for the Kallen-Lehmann representation. We briefly discuss
the implications of this violation of the 
positivity of the spectral density and
also discuss possible origin of such a violation in these models. 
Put simply this means that such models are not physical descriptions of the
hadron spectrum. 
\end{abstract} 

\begin{section}*{}
\par The correct and complete theory describing the confinement of quarks 
and gluons into hadrons, and explaining the hadronic properties 
from first principles 
was, and still is, one of the major challenges. Though there exist many
phenomenological models which seem to describe the low energy data reasonably
well,
a complete theoretical understanding is still lacking. The holographic
principle or the so called AdS/CFT correspondence \cite{adscft}
has opened a new window to
look into the problem from a completely new and different perspective. Very
loosely speaking, the correspondence is a conjecture of the 
equivalence of the generating functional of 
(large $N$ limit of) certain conformal field theory (CFT) in $d$-dimensions
and the $(d+1)$-dimensional 
supergravity (string theory in a particular limit to be more
precise, though we will use the word supergravity throughout the note) 
effective action evaluated at the boundary. The boundary value of the bulk
fields is supposed to play the role of the source in the CFT generating
functional. If we generically denote the $(d+1)$-dimensional fields as $\phi$,
then the correspondence can be schematically summarized as
\bea
Z_{AdS}[\phi_0] &=& \int_{\phi_0}~{\mathcal{D}}\phi~exp(-I[\phi]) \\ \nonumber
&\equiv& Z_{CFT}[J=\phi_0] = \langle exp(\int~d^dx {\mathcal{O}}\phi_0)\rangle
\eea
where $\phi_0$ is the boundary value of the field $\phi$ which acts as a
source corresponding to the operator ${\mathcal{O}}$ 
in the CFT generating functional. For the $(d+1)$-dimensional AdS space, 
the line element is 
\be
ds^2 = \frac{L^2}{z^2}(dx_{\mu}dx^{\mu} + dz^2)
\ee
where $x^{\mu}$ are the the $d$-dimensional coordinates and $z$ parameterizes
the extra direction, with $z=0$ describing the ultra-violet 
(UV) boundary and $L$ denotes the
radius of curvature, which we set to unity.

\par What it
means in practical terms is that a perturbative calculation on the
supergravity side can be translated into a non-perturbative result in the
gauge theory sector and vice-versa. If this be so, it simply implies
that the seemingly 
impossible task of getting the genuine non-perturbative results and
calculating the non-perturbative quantities within a gauge theory is no more
impossible. Instead, one can approach the problem from the dual side - where
perturbative calculations, though hard, are possible. This has motivated
considerable interest in getting the hadron spectrum and properties.
Till very recently, most of the models constructed and studied can be broadly
thought to fall under the {\em top-down} category of models 
i.e. the approach was to start from some string (or
supergravity) theory and try to obtain the low energy description by
demanding/imposing certain consistency conditions. For some of the earlier 
works, see \cite{susymodels}. Encouraged by these 
explorations, non-supersymmetric
holographic dual models for hadrons have recently been proposed \cite{brodsky},
\cite{sonetal}. Compared to previous studies, these models are
phenomenological and the approach adopted is {\it bottom-up}.
Guided by the basic ideas of the correspondence principle, 
one identifies the corresponding conserved
currents which appear in the gauge theory as a dual description to the
fields in the supergravity theory. This thus determines the (minimal) field
content of the five dimensional theory of gravity, as dictated directly by the
low energy sector. Thus the name, bottom-up. The models are still 
at the stage of being called {\em toy models}
 and they employ a very small sub-set of the possible field content.
One can then go ahead and calculate various $n$-point Green functions
(or correlators) of the fields in the 5-dimensional gravity theory. The
correspondence relates such a calculation to the correlators of conserved
currents in a suitable gauge theory. Using only a minimal
sub-set of fields, the authors have shown that the models are quite robust and
predictive to within $10\%$ accuracy. 
To capture the essentials of chiral symmetry
 breaking, some specific boundary conditions are imposed on the fields and
 their derivatives on the so called ``infra-red (IR) boundary'' 
($z_{IR} >> z_{UV}$). The
 theory is conformal only close to the UV boundary. As one moves away
 from this UV boundary, the theory is no more conformal. 
In a complete microscopic
 description, this should be incorporated by appropriately modifying the
 geometry which was AdS to start with. In the absence of a complete
 description, this is done by putting certain artificial and ad-hoc
boundary conditions and one hopes to capture the essential 
and broad features.    

\par In this Letter we take a closer look at these models and
explore them in more detail. Given the robustness of the predictions
within these models, it is tempting to investigate how far can they go in
describing the experimental observations. The two point functions are the
simplest and most straightforward to be evaluated within the 5D gravity
theory. Their importance is not just being the simplest objects calculable, but
lies in the fact that they offer a dual description to the current-current two
point correlator ,$\langle j_{\mu}j_{\nu}\rangle$, in the gauge theory. 
This correlator shows up in $e^+e^-$ annihilating into
hadrons, $\tau$ hadronic decays, hadronic contribution to $(g-2)_{\mu}$
 and many other places. Moreover, the
current-current correlator is used to extract the masses and decay constants
of the mesons in the large $N_c$ limit of QCD. Furthermore, it is known that 
the masses and decay constants obey certain sum rules, namely the Weinberg
sum rules and generalizations \cite{sumrules}. 
It is thus important to have a precise and
accurate theoretical description of the two point current correlation
function.

\par Before proceeding to explore the two point function in more detail, let
us briefly mention the field content and other details of the {\em toy model}
\cite{sonetal} 
that we will be working with. We follow \cite{sonetal} for the notation and
general setup of the model.
The field content of the 5D theory is dictated
by the operators in the 4D QCD. In principle, there should be an infinite
number of fields corresponding to an 
infinite number of gauge invariant operators
in QCD. However, for the purpose of chiral symmetry breaking and its essential
consequences, it suffices to look at a very small number of operators and
therefore a small number of fields in the 5D theory. As is known, the
chiral dynamics is quite effectively described by an $SU(N_f)_L\times
SU(N_f)_R$ theory. We therefore have the following three operators that  are
most crucial to this effect (corresponding 5D fields are written in front of
each of them): 
\[
\bar{q}_L(x)\gamma^{\mu}T^aq_L(x) \stackrel{dual}\longrightarrow 
A^{a\mu}_L(x,z)
\] 
\be
\bar{q}_R(x)\gamma^{\mu}T^aq_R(x) \stackrel{dual}\longrightarrow 
A^{a\mu}_R(x,z)
\ee
\[
\bar{q}_L^i(x)q_R^j(x) \stackrel{dual}\longrightarrow (2/z)X^{ij}(x,z)
\]
where $T^a$ are the group generators and $i,j$ are the flavour
indices. Restricting to two flavours implies that $i,j=1,2$ and the
$T^a$'s are the three Pauli matrices. We are thus looking at a chiral
$SU(2)_L\times SU(2)_R$ theory. With this minimal field content and ignoring
any interactions for the time being, the 5D action is
\be
S_{(5)} = \int d^4x dz \sqrt{g}~Tr\left[-\frac{1}{4g_5^2}(F_L^2 + F_R^2) + 
\vert DX\vert^2 - m^2X^2\right]
\ee
where $F^{AB}_{L,R}$ denotes the field strengths for the left and the right
gauge fields and $D_{A}$ denotes the covariant derivative. 

\par At the IR boundary, some boundary conditions are imposed on the
fields. Moreover, the reason behind introducing the IR boundary in the theory
is to parameterize the effect of chiral symmetry breaking in an effective
fashion. Else, one would be forced to examine how the 
geometry of the bulk changes
away from the UV boundary and what should be the metric far away that will
suitably describe chiral symmetry breaking in the 4D gauge theory. The
introduction of the IR boundary and suitable boundary terms in a 
phenomenological way
takes care of this aspect. For the gauge fields, we impose the boundary
conditions: $F_{L,R}^{\mu z}=0$ and choose to work in the gauge $A_z=0$. For
the field $X$, the expectation value at the UV boundary is the quark mass
matrix and the quark condensate will effectively fix the other constant
(at the IR boundary) in
the solution to the equation of motion for field $X$. In principle, the UV
boundary corresponds to $z=0$. However, in practice, the boundary conditions
are specified at $z=z_{UV}$ and finally the limit $z_{UV}\to 0$ is taken.

\par Define the vector and axial-vector gauge fields as
appropriate linear combinations of $A_L$ and $A_R$. Let us focus on the vector
gauge field $V_A = \frac{1}{2}((A_L)_A + (A_R)_A)$ and let $\tilde{V}_A(q,z)$
denote the Fourier transform of the vector field with respect to the 4D
coordinates. We have suppressed the group index for convenience.
In the $V_z=0$ gauge, the
linearized equation of motion for the transverse part reads 
\be  
\partial_z\left(\frac{1}{z}\partial_z\tilde{V}_{\mu}(q,z)\right) +
\frac{q^2}{z}\tilde{V}_{\mu}(q,z) = 0
\ee
The solution is a linear combination of the Bessel functions and can be
written as
\be
\tilde{V}_{\mu}(q,z) = C_{\mu}(q) qz[b{\mathcal{J}}_1(qz) + 
{\mathcal{Y}}_1(qz)]
\ee
The boundary condition at the IR boundary, namely
$\partial_zV_{\mu}\vert_{z=z_{IR}}=0$, fixes 
the relative constant between the two terms to be
\be
b \approx -\frac{{\mathcal{Y}}_0(qz_{IR})}{{\mathcal{J}}_0(qz_{IR})}
\label{const} 
\ee

\par We can now use the correspondence principle to interpret the above
solution at $z=z_{UV}$ as the source for a current. For the two point function
we get \cite{nima},
\be
\langle j_{\mu}(0)j_{\nu}(q)\rangle = \left(\eta_{\mu\nu} -
\frac{q_{\mu}q_{\nu}}{q^2}\right)\Pi(q)
\ee
where 
\be
\Pi(q)\vert_{z=z_{UV}} =
\left(\frac{1}{g_5^2}\right)\frac{q}{z_{UV}}
\left[\frac{{\mathcal{Y}}_0(qz_{UV}) + b {\mathcal{J}}_0(qz_{UV})}
{{\mathcal{Y}}_1(qz_{UV}) + b {\mathcal{J}}_1(qz_{UV})}\right] \label{cor1}
\ee
This expression can be expanded about $z_{UV}=0$ and we retain the leading
terms, which leads to
\be
\Pi(q)\vert_{z=z_{UV}} \approx \left(\frac{1}{g_5^2}\right)
q^2\left[\log(qz_{UV}/2) + \gamma_E + \frac{\pi}{2}b\right] \label{cor2}
\ee
The logarithmic term may be identified as the contribution to the
current-current correlator arising from the lowest order quark bubble
diagram. Comparing this expression with the one obtained within
QCD with $N_c$ colours fixes the 5D gauge coupling, $g_5$, in terms of $N_c$.

\par The next task is to estimate the hadron masses and decay constants. This
is easily done by comparing the two point function obtained above with the
corresponding expression that one obtains in the large $N_c$ chiral
theories. 
In such a theory, the two point function is expressed as a sum over
narrow resonances. The poles of the two point function yield hadron masses and
the residues at each pole are the decay constants. The correlator thus has the
following form (the $i\epsilon$ is implicit) 
\be
\Pi(q^2) = q^2\sum_n \frac{F_n^2}{q^2 - M_n^2}
\ee
In particular, the above form and the fact that we have approximated the
correlator as a sum over narrow resonances imply that the spectral density is
a {\em comb of delta functions} peaking at the hadron masses. The spectral
density is nothing but the imaginary part of $\Pi(q^2)$.
Also, let us recall that the Kallen-Lehmann representation for the two point
function in a generic field theory implies that the 
spectral density is a positive quantity
\cite{itzykson}. This is
an important property that it must satisfy and we'll see that this property
plays an important role in our analysis.

\par To estimate the hadron masses, we look for the poles of the
two point function. We can use either Eq(\ref{cor1}) or Eq(\ref{cor2}) for
this purpose. The poles are given by the zeros of
${\mathcal{J}}_0(qz_{IR})$. Then using the experimental value for 
$m_{\rho}$ as an input fixes $z_{IR}$. The
decay constants are the residues at these poles. 

\par The estimated numbers for the
masses and decay constants \cite{sonetal} 
compare well with the experimental values and we
get the impression that the model is very robust and quite close to
reality. However, as we'll see below, this is not the real and complete
picture. We take a
closer look at the two point function. In particular, we are interested in the
corresponding spectral density, $\rho(q^2)$. Both the vector and axial-vector
spectral densities are measured to very good accuracy, for example,
in the hadronic $\tau$
decays and are directly related to the decay width by optical theorem
\cite{expt}. The data clearly shows the $\rho$ and $a_1$ resonance peaks 
and supports the theoretical expectation of the spectral
densities to be positive functions. We use Eq(\ref{cor2}) (or equivalently
Eq(\ref{cor1})) to extract the imaginary part, and therefore the spectral
density. Using the usual $i\epsilon$ prescription, it is easy to obtain the
imaginary part which is nothing but the following
\be
\rho(q) = {\it Im}\left[
  \frac{{\mathcal{Y}}_0(qz_{IR})}{{\mathcal{J}}_0((q+i\epsilon)z_{IR})}\right] 
\ee 

\par We would like to study the above spectral density in more detail
and convince ourselves that it satisfies all the field theoretic properties,
like the positivity condition, before we can proceed further and calculate
masses and decay constants from the two point correlation function. To this
end, we consider the following function
\be 
f(x) = \frac{{\mathcal{Y}}_0(x)}
{{\mathcal{J}}_0(x+i\epsilon)} = {\it Re}[f(x)] + i {\it Im}[f(x)] 
\ee 
Clearly, the imaginary part of the function is nothing but the spectral 
density itself rewritten in terms of a different variable. In Figure 1, we have
plotted the imaginary part of the function, ${\it Im}[f(x)]$.\footnote{
A quick way to see that the imaginary part will have a profile very similar to
a sum of delta functions is to look at the series expansion, 
${\mathcal{J}}_0(x)
= 1 - \frac{x^2}{4} + \frac{x^4}{64} - \frac{x^6}{2304} + ....$. 
The $i\epsilon$ prescription can now be used to obtain the imaginary
part. For plotting the graph, we have chosen 
$\epsilon = 10^{-7}$ and rescaled
the y-axis. It should be borne in mind that any other small value of
$\epsilon$ is equally good and rescaling simply helps in having an enlarged
picture and does not change the shape and nature of the curve.}
A quick look at
the plot gives the impression that the spectral density is indeed a comb of 
delta functions, as expected and desired. The zeros of the Bessel function
${\mathcal{J}}_0(x)$ are the positions of the resonance masses
 and the residues at
these values will correspond to the decay constants of the mesons.
Let us now take a more closer look at the function itself. 
The function under investigation is a ratio of two Bessel functions. Further, 
the Bessel functions are known to have an oscillatory behaviour, with 
${\mathcal{J}}_0(x)$ and ${\mathcal{Y}}_0(x)$ having opposite behaviour with
respect to each other. Figure 2 is a plot of the two Bessel functions, clearly
showing these features.

\par The relevant quantity (the imaginary part of the function $f(x)$ which is
the spectral density) is a ratio of these two
different Bessel functions and because of the features mentioned above, 
${\mathcal{Y}}_0(x)$ can cross the real axis between two zeros of 
${\mathcal{J}}_0(x)$, thereby yielding negative values for the imaginary part
of the function, and therefore the spectral density. To further substantiate
our claim, we look at the behaviour of the imaginary part of the function 
$f(x)$ more closely. In Figure 3, we plot $Im[f(x)]$ for various 
smaller intervals of $x$ and show that it indeed acquires negative values. The
reason that this feature is not evident in Figure 1 is due to the very large
values that the function acquires close to the resonances. However, when we
look at the behaviour in regions slightly away from the resonances, the
negative values show up, as shown in Figure 3.
\vskip 1.0cm
\begin{figure}[ht]
\vspace*{-1cm}
\centerline{
\epsfxsize=9.0cm\epsfysize=5cm
                      \epsfbox{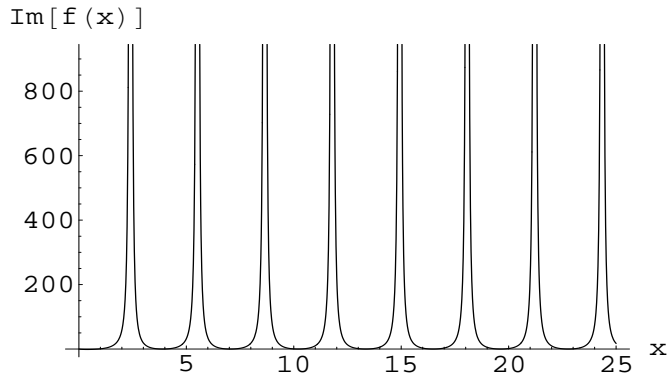}}
\caption{{\it Coarse-grain} view of ${\it Im}[f(x)]$ as a 
function of the argument, $x$}
\end{figure}
\begin{figure}[ht]
\centerline{
\epsfxsize=5.0cm\epsfysize=4cm
                      \epsfbox{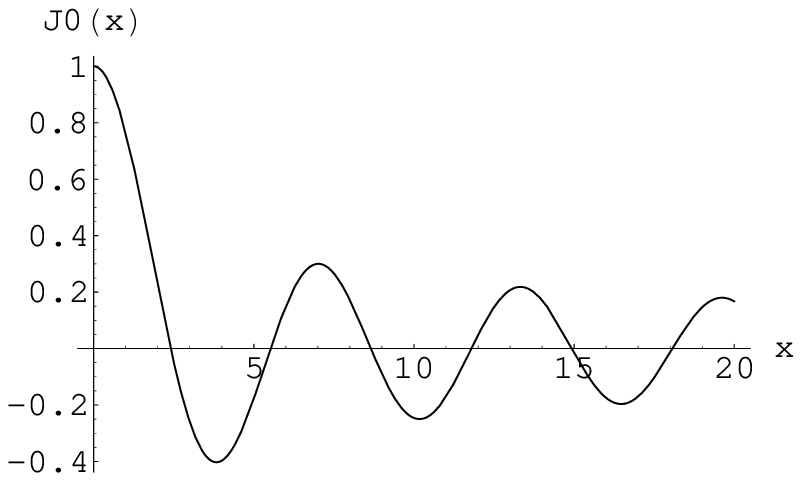}
\hskip 1cm
\epsfxsize=5.0cm\epsfysize=4cm
                      \epsfbox{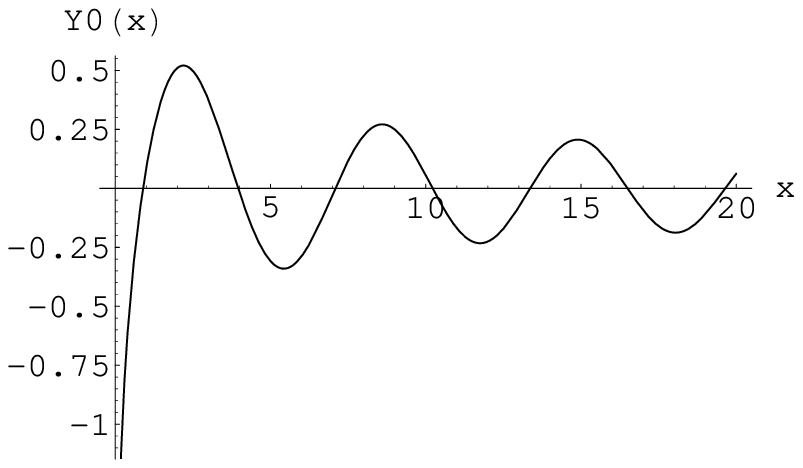}}
\caption{The two Bessel functions: ${\mathcal{J}}_0(x)$ (left) 
and ${\mathcal{Y}}_0(x)$ (right)}
\end{figure}
\begin{figure}[ht]
\centerline{
\epsfxsize=3.5cm\epsfysize=4cm
                      \epsfbox{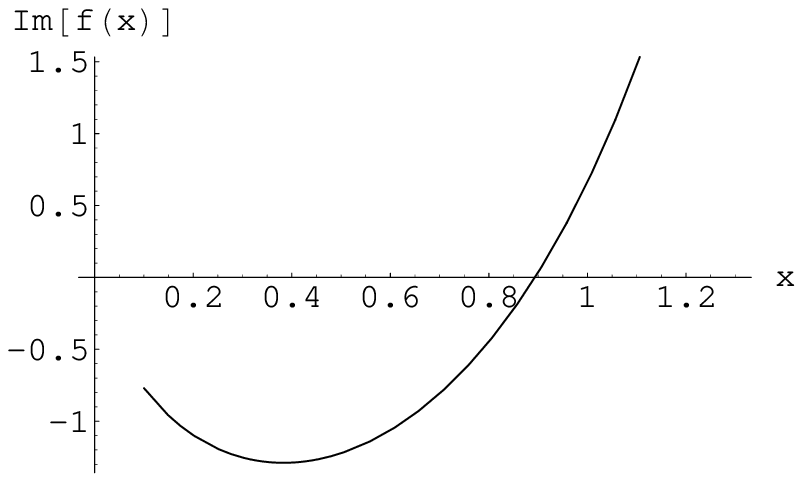}
\hskip 0.5cm
\epsfxsize=3.5cm\epsfysize=4cm
                      \epsfbox{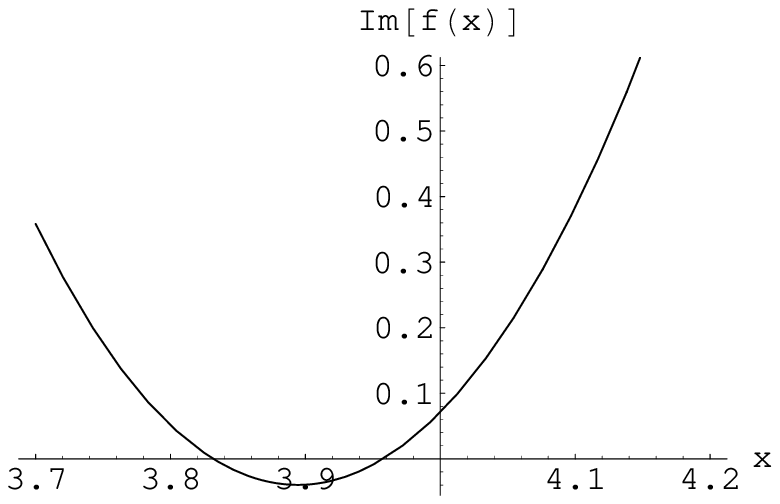}
\hskip 0.5cm
\epsfxsize=3.5cm\epsfysize=4cm
                      \epsfbox{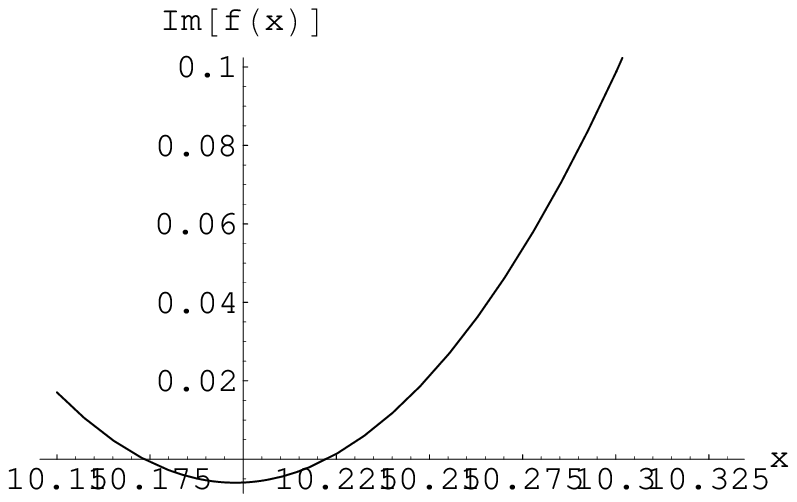}}
\caption{${\it Im}[f(x)]$ plotted in different intervals}
\end{figure}
\par Let us put all the individual pieces of information together to get a
final and complete picture. We have seen above that a cursory look at the
imaginary part gives the impression that it is a comb or sum of delta
functions - the position of the peaks will give the masses of the mesons and
the residues at these values will be the respective decay constants. It is
quite clear from the Figure 1 that the residues at the peak positions of
the imaginary part are indeed positive. It is this {\it coarse-grain} picture
that leads us to believe that all is well with the model and the predicted 
numerical values for the meson masses and decay constants are in 
agreement with the experimental values to within $10-20\%$. 
However, as is clear from Figure 3, the imaginary part of
the function (the spectral density) acquires negative values. 
Also, it is clear from the figure that
we encounter negative values in a somewhat periodic manner and that
the magnitude of the negative values attained diminishes as we go to larger
values of the argument. This feature is also not hard to expect and
understand. The ``somewhat periodic'' appearance of the 
negative values is due to
the oscillatory behaviour of Bessel functions (or their combination) while
from the behaviour of the Bessel functions, it is very evident that 
the amplitude keeps on decreasing for larger and larger values of the argument.

\par It is now straightforward to convince ourselves that the spectral density
obtained within these models is not positive semi-definite. 
However, as we have discussed earlier, the spectral density
is a positive quantity and has been very well measured in experiments, in full
conformity with our theoretical expectations. This is the most
important observation and result of this study. It may be worthwhile to
briefly comment on the results/findings of the models
\cite{brodsky,sonetal} in the light of this observation. As has been mentioned
above, if we just content ourselves with the positions of the resonances
(meson masses) and evaluate the residues at these positions (decay constants), 
we'll obtain positive values for them. This is essentially what is done in
\cite{brodsky,sonetal}. It is only when we take a closer look at the function
under consideration that we are led to the observation that the imaginary part
of the function, which is supposed to be positive always, acquires negative
values.

\par The above observation regarding the spectral density becoming negative is
a serious issue of concern. As was mentioned, the spectral density is directly
related to the hadronic $\tau$ decay width by the optical theorem. 
A negative spectral density will simply mean that the decay width is 
becoming negative - something that is clearly unphysical and of course
unobserved. Translated back, this observation has to say something very
important about the model itself. In this form, the model does not lead to
physical predictions and therefore can not be trusted. Similar conclusions can
be reached at from the data on $e^+e^-$ annihilating into hadrons or the
hadronic contributions to $(g-2)_{\mu}$. In each case the spectral density is
related to a positive physical observable like cross-section or decay rate.

\par We make a brief attempt to discuss the possible origin of such a problem
in these models. 
Recall that to capture the essential features of chiral symmetry
breaking, an ad-hoc and artificial infra-red boundary was introduced in the
theory and certain specific boundary conditions specified on it. This approach
is completely phenomenological and though, intuitively may seem well motivated
and correct, by itself, does not guarantee that the results will be unitary
and physical. As was pointed out initially, in a complete microscopic
description, the geometry of the space-time should be appropriately modified
and this should be consistently done so that away from the AdS boundary
($z=0$), 
the model incorporates the correct pattern for chiral dynamics. 
However, in the case at hand, this
was avoided by invoking the artificial IR boundary. In our opinion, this
itself is the root cause of the problem. The reason is as follows. Both the
masses and the decay constants are obtained by essentially looking at the
constant $''b''$ that appears in the solution to the equation of motion for the
gauge field. This constant is fixed in the present scenario by requiring the
derivative of the solution to vanish at the artificial IR boundary, thus
yielding a ratio of two Bessel functions. It is this combination of Bessel
functions that leads to the trouble. We may be led to speculate that 
if, instead of the approximate form for
the relative constant $b$ in Eq(\ref{const}), we had used the full expression,
we would have bypassed the problem. It is again easy and straightforward to
convince ourselves that this is not the case as the new form is also a
combination of some other Bessel functions.
If however, we can model the chiral
symmetry breaking pattern by continuously changing the geometry in the bulk,
this problem can possibly be avoided.
We would like to point out that this is a common problem in all the
models that invoke an IR boundary condition to model the chiral symmetry
breaking. We restricted ourselves to the vector sector of the theory but the
same arguments and conclusions apply to the axial vector sector as well. 
One can also
check that the Weinberg sum rules are not satisfied in these models and there
is no explanation of the (approximate) $\rho$ meson dominance that is observed
in nature.

\par We conclude by summarizing our main observation and some of its
consequences. We have investigated the profile of the spectral density in the
recently proposed holographic models of QCD or chiral symmetry breaking in
five dimensions \cite{brodsky,sonetal}. 
The models seem to be quite robust and naively taken, seem to
be predictive to within $10\%$ accuracy. However, a closer look reveals that
the spectral density, extracted from the two point correlation function, keeps
acquiring negative values. This is in contrast to
the positive behaviour of the spectral density expected from very general field
theory arguments. Moreover, the vector and the axial-vector spectral densities
are directly measured, for example, 
in the corresponding hadronic $\tau$ decay modes. A
negative spectral density implies a negative decay rate, in clear violation
with unitarity and optical theorem and also with the observed data. This
simply implies that the proposed models, though seem remarkably predictive,
do not satisfy some of the basic field theoretic requirements and therefore
can not be trusted. Also, in this form they can not be seriously taken to be
models describing the physical hadron spectrum. Similar arguments will hold
for any other model that violates the positivity condition for the spectral
density and/or is in conflict with generic field theoretic expectations. 
As pointed out, the root cause in the present case
is the way chiral symmetry breaking has been modeled by introducing an
ad-hoc IR boundary.\footnote{This feature is expected to be present in any
  model that invokes an artificial IR boundary condition to mimic chiral
  symmetry breaking. We would also like to mention that the observations of
  the present study need not apply to the works listed in \cite{susymodels}.}
If on the other hand, this is done in a more consistent
manner by suitably modifying the bulk geometry in a continuous fashion, there
is hope to get a dual model of hadrons which avoids the above mentioned
problem. 

\vskip 1cm
{\bf Acknowledgements} I would like to thank for Somdatta Bhattacharya, 
Kazuyuki Furuuchi, Rajesh Gopakumar, Sukanta Panda, Ashoke Sen 
and K.~P.~Yogendran for discussions. I wish to thank
Rajesh Gopakumar for a careful reading of the manuscript.

\end{section}

\end{document}